\documentclass[12pt]{article}
\usepackage{a4}
\usepackage{graphicx}
\usepackage{amsmath}
\usepackage{latexsym}

\newcommand{\Ref}[1]{(\ref{#1})}
\newcommand{\gse}{ground state energy }
\newcommand{\gsep}{ground state energy. }
\newcommand{\hke}{heat kernel expansion }

\newcommand{\hkks}{heat kernel coefficients }
\newcommand{\hkksp}{heat kernel coefficients. }
\newcommand{\uv}{ultraviolet }
\newcommand{\be}{\begin{equation}}\newcommand{\ee}{\end{equation}}
\newcommand{\bea}{\begin{eqnarray}}
\newcommand{\eea}{\end{eqnarray}}
\newcommand{\beq}{\begin{eqnarray}}
\newcommand{\eeq}{\end{eqnarray}}
\newcommand{\beao}{\begin{eqnarray*}}
\newcommand{\eeao}{\end{eqnarray*}}
\newcommand{\nn}{\nonumber}
\renewcommand{\d}{{\rm d}}
\newcommand{\la}{\lambda}

\begin{document}

\title{\bf Dependence of the vacuum energy on spherically
symmetric background fields}

\author{Michael Bordag\thanks{E-mail: Michael.Bordag@itp.uni-leipzig.de}, \
Meik Hellmund\thanks{E-mail: Meik.Hellmund@itp.uni-leipzig.de}, \\
Universit{\"a}t Leipzig, Institut f{\"u}r Theoretische Physik,\\
Augustusplatz 10, 04109 Leipzig, Germany  \\
\\
Klaus Kirsten\thanks{E-mail: klaus@a13.ph.man.ac.uk},\\
The University of Manchester, Department of Theoretical Physics\\
Manchester, England} 

\maketitle

\begin{abstract}
The vacuum energy of a scalar field in a spherically symmetric
background field is considered. Based on previous work [Phys. Rev. D
{\bf 53} (1996) 5753], the numerical procedure is refined further and
applied to several examples.  We provide numerical evidence that
repulsive potentials lead to positive contributions to the vacuum
energy. Furthermore, the crucial role played by bound-states is
clearly seen.
\end{abstract}

Pacs number(s): 11.10-z; 02.30.-f; 11.10Gh

\section{Introduction}

The calculation of vacuum energies in the context of quantum field theory
under external conditions plays an important role in several areas of modern
theoretical physics. External conditions present may be classical fields (for
example gravitational \cite{Birrell:1982} and electromagnetic fields,
monopoles \cite{tHooft:1974,Polyakov:1974ek}, Sphalerons
\cite{Klinkhamer:1984di} or electroweak Skyrmions \cite{Skyrme:1962vh}), or
they may be realized through boundaries at which the quantum field has to
satisfy certain boundary conditions (see for example \cite{Elizalde:1994}).
External fields most naturally appear as classical solutions of some physical
nonlinear equations, but they may also be viewed as a localized property of
space representing a distribution of matter within which the quantum field
exists.  For external fields realized as boundaries this situation is usually
called Casimir effect. There is a number of generalizations towards weakened
boundaries, e.g., by introducing semi hard walls \cite{Actor:1995vc} or
transparent boundaries realized by delta functions \cite{Bordag:1992cm}. For
actual interest in such calculations see, e.g.,
\cite{Kim:1998wz,Scoccola:1998eq}.

A formalism how to deal with arbitrary background fields, which we use here,
has been developed in \cite{Bordag:1995jz} for a background depending on one
cartesian coordinate and in \cite{Bordag:1996fv} for a general spherically
symmetric background. However, these methods had been applied to a square
well potential only where the interesting quantities can be expressed in terms
of Bessel functions.

The aim of the present work is to show that this formalism is well suited for
arbitrarily shaped spherically symmetric background fields. We investigate in
detail different types of background fields as examples. It is observed that
the main feature of the vacuum energy is determined by the number (and depth)
of bound states.  As long as these are kept fixed the general behavior of its
energy when varying parameters is the same within the class of examples
considered. At the same time we find that  the presence of bound states
does not fix the sign of the \gsep

Similar calculations, however with a less extend of details and using other
techniques, have been performed earlier, e.g., in \cite{Baacke:1990sb}
and \cite{Farhi:1998vx}.

The paper is organized as follows. In the next section the basic notations of
the method used are explained. In the third section the 'asymptotic'
contribution is calculated. In the fourth section the numerical procedure is
developed in detail followed by a section containing the examples calculated.
The results are discussed in the conclusions.
\section{The model, its renormalization and some basic results}

We  consider the theory described by the Lagrangian
\begin{eqnarray}
L=\frac 1 2 \Phi (\Box -M^2 -\lambda \Phi ^2 )\Phi
+\frac 1 2 \varphi (\Box -m^2 -\lambda ' \Phi^2 )\varphi .\label{22.1}
\end{eqnarray}
Here the field $\Phi$ is a classical background field. By means of
\begin{eqnarray}
V(x) =\lambda ' \Phi ^2  \label{22.2}
\end{eqnarray}
it defines the potential in (\ref{22.1}) for the field $\varphi (x)$, which
should be quantized in the background of $V(x)$. 

Our interest is in the vacuum energy of the theory which is defined as 
half the sum of the one particle energies of the field $\varphi(x)$
\begin{eqnarray}\label{22.4}
E_{\varphi}[\Phi ] =\frac 1 2 \sum_{(n)} (\lambda_{(n)} ^2 +m^2 )
 ^{1/2-s} \mu^{2s} \,.
\end{eqnarray}
Here $s$ is the regularization parameter with $s\to 0$ in the end and $\mu$ is
an arbitrary mass parameter. The $\lambda_{(n)}^{2}$ are the one-particle
energy eigenvalues determined through
\begin{eqnarray}
(-\Delta +V(x))\phi_{(n)}(x) =\lambda _{(n)}^2 \phi_{(n)}(x) .\label{22.5}
\end{eqnarray}
For the moment a finite volume is assumed to have discrete eigenvalues.

The spectral function appearing in Eq. (\ref{22.4}) is related to 
the zeta function of the wave operator (\ref{22.5}),
\begin{eqnarray}
\zeta_V (s) = \sum_{(n)} (\lambda_{(n)} ^2 +m^2 )^{-s}\,,\label{zetascat}
\end{eqnarray}
by
\begin{eqnarray}
E_{\varphi } [\Phi ] =\frac 1 2 \zeta_V (s-1/2) \mu^{2s}.\label{groundzeta}
\end{eqnarray}
The \uv divergences contained in the \gse \Ref{22.4} needs to be
renormalized. The divergences are known and best expressed in terms of the
 \hkks related to the operator defined in Eq. \Ref{22.5}. Then the divergent
 part of the \gse reads
\begin{eqnarray}
E_{\varphi} ^{div} [\Phi ] &=& -\frac{m^4}{64\pi^2}\left(\frac 1 {s}
+\ln\frac
{4\mu^2}{m^2} -\frac 1 2 \right)A_{0}\nonumber\\
& &+\frac{m^2}{32\pi^2}\left(\frac 1 {s}
+\ln\frac{4\mu^2}{m^2} -1\right) A_{1}\label{endiv}\\
& &-\frac 1 {32\pi^2} \left(
\frac 1 {s} +\ln \frac{4\mu^2}{m^2}-2\right)A_{2},\nonumber
\end{eqnarray}
with the coefficients $A_{1}=-\int\d\vec{x} \ V(x)$, $A_{2}=\frac12
\int\d\vec{x} \ V(x)^{2}$ and $A_{0}$ is the spatial volume. The contribution
of $A_{0}$, because it does not depend on the background will be dropped from
here on without further comment. 

The renormalization procedure is set by the definition
\begin{eqnarray}
E_{\varphi}^{ren} =E_{\varphi} [\Phi ] -E_{\varphi}^{div}
[\Phi ]\label{renen}
\end{eqnarray}
and the corresponding counter terms can be absorbed into a redefinition  of the
parameters $M^{2}$ and $\la$  of the classical background field whose energy
reads
\begin{equation}\label{22.3}
E_{\rm class}[\Phi]= \frac12 \int
\d\vec{x}\left(\nabla\Phi\right)^{2}-\frac{M^{2}}{2\lambda'}A_{1}+
 {\la\over{\lambda'}^2}  A_{2}\,.
\end{equation}
It might be useful to note that among the divergent contributions \Ref{endiv}
to the \gse there are no terms proportional to the kinetic part in the
Lagrangian of the background $\Phi(x)$. Within the given framework this is a
simple consequence of the structure of the \hkksp Sometimes for the divergent
part of the \gse the first 2 terms of the perturbative expansion of
$E_{0}^{}[\Phi]$ with respect to powers of the background field $\Phi$ is
taken (e.g., in \cite{Farhi:1998vx}). This has the disadvantage 
of containing nonlocal
contributions (in opposite to the \hke expansion which is local) and
invites to discuss the kinetic term among the divergent ones which is in fact
unnecessary.

As the renormalization procedure is not unique, a normalization condition must
be imposed. We require
\begin{eqnarray}
\lim_{m\to\infty} E_\varphi^{ren} [\Phi ] =0\,.\label{norm}
\end{eqnarray}
For a massive field this is the natural condition requiring the energy of
the quantum fluctuation to vanish for the field becoming heavy. As discussed
in \cite{Bordag:1999vs} this delivers a unique definition of the \gse (in
opposite to the case of a massless field, see the discussion in
\cite{Bordag:1999vs} and also in \cite{Blau:1988kv}). 

To proceed we follow \cite{Bordag:1996fv} and use the representation of the
regularized \gse in terms of the Jost function $f_l(k)$ of the scattering
problem corresponding to \Ref{22.5}
\begin{eqnarray}
E_\varphi [\Phi ]  =-\frac{\cos \pi s}{ \pi}\mu^{2s} \sum_{l=0}^{\infty}(l+1/2)
\int\limits_{m}^{\infty}dk\,\, [k^2-m^2]^{\frac{1}{2}-s}~
\frac{\partial}{\partial k}\ln  f_l (ik)\,.\label{2.5}
\end{eqnarray}
We remind the definition of the so-called regular solution $\phi_{n,l}(r)$ of
the radial wave equation (see, e.g., \cite{Taylor:1972})
\begin{eqnarray}
\left[\frac{d^2 }{dr^2} -\frac{l(l+1)}{r^2} -V (r)
+k^2 \right] \phi_{n,l} (r) =0\,. \label{2.1}
\end{eqnarray}
It is the solution, which for $r\to 0$ behaves as the free
(i.e., without potential) solution. The Jost function (and its complex
conjugate) are defined as the coefficients in the asymptotic expansion of this
solution as $r\to\infty$:
\begin{eqnarray}
\phi_{l,p}(r) &\sim& \frac i 2 \left[ f_l(p) \hat h _l^- (pr) -
        f_l^*(p) \hat h _l^+ (pr) \right],\label{2.2}\\[-5pt]
&{\stackrel{r\to\infty}{}}&\nonumber
\end{eqnarray}
where $\hat h _l^- (pr)$
and $\hat h _l^+ (pr)$ are the Riccati-Hankel functions.

Instead of dealing explicitly with eigenvalues, the vacuum energy is expressed
in terms of the Jost function.  Although representation (\ref{2.5}) does not
reveal explicitly the dependence of the energy on the bound states, in
\cite{Bordag:1996fv} it was shown that their contribution is correctly
included. This, and the non oscillating behavior of the Jost function on the
imaginary axis (cf.  \Ref{2.5}) gives this representation calculational
advantages with respect to alternative representation (e.g. integrating the
scattering phases over the real $k$-axis).

To proceed we have to perform the analytic continuation in $s$ to $s=0$ in
representation \Ref{renen} of $E_{\varphi}^{\rm ren}$ using \Ref{2.5} for
$E_{\varphi}[\Phi]$. Although we known that the result is finite, this task
requires some work. The point is that one cannot put $s=0$ under the sign of
the sum and the integral in \Ref{2.5} being inserted into Eq. \Ref{renen}. The
reason is the lack of convergence for $k$ and $l$ large. Therefor we use the
uniform asymptotic expansion $\ln f_{k}^{\rm asym}(ik)$ of the logarithm of
the Jost function for both, $k$ and $l$ large. We subtract and add it to the
integrand. As a result we get the sum of two parts
\begin{eqnarray}
E_\varphi^{\rm ren} [\Phi ]=E_f[\Phi ]+E_{as} [\Phi ] \label{3.3a}
\end{eqnarray}
with
\begin{eqnarray}
 E_f [\Phi ]=-\frac{1}{\pi} 
\sum_{l=0}^{\infty}(l+1/2) \int\limits_{m}^{\infty}dk\,\,
 [k^2-m^2]^{\frac{1}{2}}\frac{\partial}{\partial k}
[\ln f_l (ik) - \ln f_l ^{asym}
(ik) ]
\label{ns}
\end{eqnarray}
and
\begin{eqnarray}
 E_{as} [\Phi ]=-\frac{\cos (\pi s)}{\pi} \mu^{2s}
\sum_{l=0}^{\infty}(l+1/2) \int\limits_{m}^{\infty}dk\,\,
[k^2-m^2]^{\frac{1}{2}-s}\frac{\partial}{\partial k}
\ln f_l ^{asym} (ik) - \ E_{\varphi}^{\rm div}\,.\label{as}
\end{eqnarray}
The first contribution, $E_f [\Phi ]$ is called the {\it finite} part. There
the regularization parameter $s$ has been put to zero directly because the sum
and the integral are convergent due to the subtraction. The second
contribution, $E_{as} [\Phi ]$, is called the {\it asymptotic} contribution. It
has, as we know, a finite limit for $s\to 0$ which will be determined below.
The asymptotic expansion $\ln f_l^{asym}(ik)$ for $\nu\to\infty$,
$k\to\infty$, $\nu/k$ fixed, was determined in \cite{Bordag:1996fv} up to the
required order of $\nu^{-3}$. It reads
\begin{eqnarray}
\ln f_l^{asym}(ik)&=&
 \frac 1 {2\nu} \int\limits_0^{\infty}dr\,\,
\frac{r\,V(r)}{\left[1+\left(\frac {kr} {\nu}\right)^2 \right]^{1/2}}
 \nonumber\\
& &+\frac 1 {16\nu^3}\int\limits_0^{\infty}dr\,\,
\frac{r\,V(r)}{ \left[1+\left(\frac {kr} {\nu}\right)^2
\right] ^{3/2}}\left[1 -
 \frac 6
  {\left[1+\left(\frac {kr} {\nu}\right)^2 \right] }
+ \frac 5  {\left[1+\left(\frac {kr} {\nu}\right)^2 \right]^{2}}\right]
\nonumber\\
& & -\frac 1 {8\nu^3}\int\limits_0^{\infty}dr\,\,\frac{r^3\, V^2 (r)}
{ \left[1+\left(\frac {kr} {\nu}\right)^2 \right]^{3/2}},
\label{3.3b}
\end{eqnarray}
with $\nu = l+1/2$.

\section{The  asymptotic contribution}

First of all the $k$-integrals in (\ref{as}) can be done explicitly using
\bea
&&{\int\limits_{m}^{\infty}dk\,\,
[k^2-m^2]^{\frac{1}{2}-s}\frac{\partial}{\partial k}}\left[1+
\left(\frac {kr} {\nu}\right)^2 \right] ^{-\frac{n}{2}}\nn \\
&&=
 -\frac{\Gamma (s+\frac{n-1}{2})
\Gamma (\frac{3}{2}-s)}{ \Gamma (n/2)}
 \frac{\left(\frac{\nu}{mr}\right)^n
m^{1-2s}}{\left(1+
\left(\frac{\nu}{mr}\right)^2\right)^{s+\frac{n-1}{2}}}.\label{kint}
\eea
This naturally leads to functions of the type
\begin{eqnarray}
f(s;a;b) =\sum_{\nu = 1/2,3/2,\ldots} ^{\infty}
\nu ^a \left( 1+ \left( \frac{\nu}
{mr } \right ) ^2 \right) ^{-s-b}, \label{funf}
\end{eqnarray}
which we need around $s=-1/2$. The way to deal with this functions
has been developed
in \cite{Elizalde:1997hx} and for completeness the relevant results
are provided in the Appendix, (\ref{anhfun}). Using these, it is a simple
exercise to find
\begin{eqnarray}\label{numasym}
&&E_{as}^{ren}[\Phi ] =
-\frac 1 {8\pi} \int_0^\infty dr \,\, r^2 V^2 (r) \ln (mr) \\
&&-\frac 1 {2\pi} \int_0^\infty dr \,\, V(r)
\int_0^{\infty} d\nu\;
\frac{\nu} {1+e^{2\pi \nu}}
\ln |\nu^2 -(mr)^2| \nonumber\\
&&-\frac 1 {8\pi} \int_0^\infty dr \,\,
\left[r^2 V^2 (r) -\frac 1 2 V(r)\right]
\int_0^\infty d\nu\, \left( \frac{d}{d\nu}\, \frac{1}{
1+e^{2\pi \nu}} \right)\, \ln \left| \nu^2 -(mr)^2 \right|  \nonumber \\
&&-\frac 1 {8\pi} \int_0^\infty dr \,\, V(r)
\int_0^\infty d\nu\,
\left[ \frac{d}{d\nu} \left( \frac{1}{\nu} \frac{d}{d\nu} \frac{\nu^2}{
1+e^{2\pi \nu}}\right)  \right]\, \ln \left| \nu^2 - (mr)^2 \right|\nonumber\\
&&+\frac 1 {48\pi} \int_0^\infty dr \,\, V(r)
\int_0^\infty d\nu\,
\left[ \frac{d}{d\nu} \left( \frac{1}{\nu} \frac{d}{d\nu}
 \frac{1}{\nu} \frac{d}{d\nu} \frac{\nu^4}{
1+e^{2\pi \nu}}\right)  \right]\, \ln \left| \nu^2 - (mr)^2 \right|,\nn
\end{eqnarray}
suitable for numerical evaluation.

\section{Numerical determination of the Jost function and the
integration procedure}

Under the assumption that the potential has compact support, let's say
$V(r)=0$ for $r\geq R$, the regular solution may be written as
\begin{eqnarray}
\phi_{l,p}(r) =u_{l,p} (r) \Theta (R-r) +\frac i 2
\left[ f_l(p) \hat h _l^- (pr) - f_l^*(p)
 \hat h _l^+ (pr) \right] \Theta (r-R).\label{4.1}
\end{eqnarray}
This solution and its derivative must be continuous at $r=R$. So the matching
conditions are
\begin{eqnarray}
 u_{l,p} (R) &=& \frac i 2 \left[ f_l(p) \hat h _l^- (pR) - f_l^*(p)
\hat h _l^+
 (pR)\right],\nonumber\\
 u_{l,p}' (R) &=& \frac i 2 p \left[ f_l(p)  {\hat h}_l^{- \prime}
 (pR) -  f_l^*(p)  {\hat h} _l^{+ \prime} (pR) \right].\nonumber
\end{eqnarray}
From these conditions we get the Jost function in terms of the
regular solutions,
\begin{eqnarray}
 f_l(p) = -\frac 1 p \left( p u_{l,p} (R)
{\hat h} _l^{+ \prime} (pR) - u_{l,p}'
   (R)
 \hat h _l^+ (pR) \right),\label{4.2}
\end{eqnarray}
where we used that the Wronskian determinant of $\hat{h}_l^{\pm}$ is $2i$.

The assumption of the potential $V(r)$ having a finite support is natural
within the framework of a numerical calculation. Anyway the potential is
assumed to vanish for $r\to\infty$. So, after some value of the radius it can
be viewed as zero within the required precision.  This simplifies the
numerical procedure considerably.  Another situation would occur for a Coulomb
like potential which we do not consider in this paper.

In order to determine a unique solution of the differential equation
(\ref{2.1}), we need to pose the initial value problem corresponding to
$\phi_{l,p}(r)$ being the regular solution. We have for $r\to 0$
\begin{eqnarray}
\phi_{l,p}(r) = u_{l,p} (r) \sim \hat j _l (pr)
\sim \frac{\sqrt{\pi}}{\Gamma (l+3/2)} \left( \frac z 2 \right) ^{l+1} .
\nonumber
\end{eqnarray}
We change to the new function
\begin{eqnarray}
u_{l,p} (r) = \frac{\sqrt{\pi}}{\Gamma (l+3/2)} \left( \frac{pr}
2 \right) ^{l+1} g_{l,p} (r) \nonumber
\end{eqnarray}
with the initial value $g_{l,p} (0) =1$. The differential equation for
$g_{l,p}(r)$ reads
\begin{eqnarray}
\left\{ \frac{d^2}{dr^2} +2\frac{l+1} r \frac d {dr} -V(r) +p^2\right\}
g_{l,p} (r) =0 , \nonumber
\end{eqnarray}
or, going to the imaginary $p$-axis and writing it as a first order
differential equation with $(\partial / \partial r) g_{l,ip} (r)
= v_{l,ip } (r)$,
\begin{eqnarray}
\frac d {dr} \left(
\begin{array}{c}
   g_{l,ip} (r) \\
   v_{l,ip} (r)  \end{array} \right) =
\left( \begin{array}{cc}
  0 & 1 \\
 V(r) +p^2 & -\frac 2 r (l+1)
  \end{array} \right)
\left(
\begin{array}{c}
   g_{l,ip} (r) \\
   v_{l,ip} (r)  \end{array} \right).\label{firstord}
\end{eqnarray}
To fix the solution uniquely we only need to fix $v_{l,ip}(0)$. A power
series ansatz for $g_{l,ip} (r)$ about $r=0$ shows that for $V(r) =
{\cal O} (r^{-1+\epsilon})$ the condition reads $v_{l,ip}(0) =0$. With this
unique solution of (\ref{firstord}), the Jost function takes the form
\begin{eqnarray}
f_l(ip) = \frac 2 {\Gamma (l+3/2)} \left( \frac{pR} 2 \right)
^{l+3/2} \left\{ g_{l,ip} (R) K_{l+3/2} (pR) +\frac 1 p g'_{l,ip} (R)
 K_{l+1/2} (pR) \right\}.\label{endjost}
\end{eqnarray}
Finally, doing a partial integration and the substitution $q=\sqrt{k^2-m^2}$
in Eq. (\ref{as}), the starting point for the numerical evaluation used for
$E_f$ is
\begin{eqnarray}
E_f = \frac 1 \pi \sum_{l=0}^\infty (l+1/2)
\int_0^\infty dq \,\, \left[ \ln f_l (i\sqrt{q^2+m^2}) -\ln f_l^{asym}
(i\sqrt{q^2+m^2}) \right] \,,\label{efinfin}
\end{eqnarray}
where $\ln f^{\rm asym}_{l}$ is given by Eq. \Ref{3.3b}.

\section{Examples}

The method described above works for radially symmetric potentials $\Phi(r)$
having compact support $\Phi(r\geq R)=0$. In order to have a finite
classical energy (Eq.~\ref{22.3}) we also demand
$\Phi^\prime (r=0) = \Phi^\prime (r=R)=0$.

In the following we use dimensionless quantities
\begin{eqnarray}
 \epsilon &=& ER,\\  \mu &=& mR,\nonumber \\ \rho&=&\frac{r}{R},\nonumber\\
 V(r) &=& \lambda^\prime \Phi^2(r)\nonumber \\
&=&
\frac{1}{R^2} v(\rho) =
\frac{\lambda^\prime}{R^2} \varphi^2(\rho) \nonumber
\end{eqnarray}
such that $\epsilon = \epsilon[\varphi, \mu]$. Therefore, large $\mu$
corresponds to large $R$ and to large $m$, respectively.

We studied two types of potentials. Type A is lump-like concentrated around
$r=0$ whereas type B has its maximum at $R/2$, acting like a spherical wall:
\begin{eqnarray}
  \label{eq:pot}
  \varphi_A(\rho) &=& \frac{a(1-\rho^2)^2}{a+\rho^2}\,,\\
  \varphi_B(\rho) &=& \frac{16 a \rho^2 (1-\rho)^2}{a+(1-2\rho)^2}\,.\nonumber
\end{eqnarray}
The parameter $a$ allows to vary the shape of the potentials, as
indicated in the insets of the figures.

Our numerical implementation of Eq.~(\ref{efinfin}) showed that the sum over
angular momenta converges quite fast, usually after 10--50 terms. Hereby,
the numerical $q$-integration procedure has to solve the system of ordinary
differential equations    Eq.~(\ref{firstord}) for every evaluation of the
Jost function $f_l(ip)$.

As a check, we compared our results with the asymptotic
formulae
\begin{equation}
  \label{eq:asy}
  \epsilon \sim
  \begin{cases}
    \frac{1}{16\pi^{2}}A_{2}\log\frac 1 \mu  \,&
 \text{for $\mu\to0$}\\
    \frac{-1}{32\pi^{2}}\frac{A_{3}}{\mu^{2}}&
       \text{for $\mu\to\infty\,, $}
  \end{cases}
\end{equation}
where $A_{2}=\frac12\int\d\vec{x}V(r)^{2}$ is the second heat kernel
coefficient and \\ $A_{3}=\frac16\int\d\vec{x}\left( \frac12 V(r)\Delta
  V(r)- V(r)^3\right))$ is the third one. These relations follow from
scaling arguments together with the well known property of the \hke being the
asymptotic expansion for large mass.

\begin{figure}[!htbp]
  \begin{minipage}[t]{18cm}
\hskip-1.5cm
\includegraphics[scale=0.9,angle=-90]{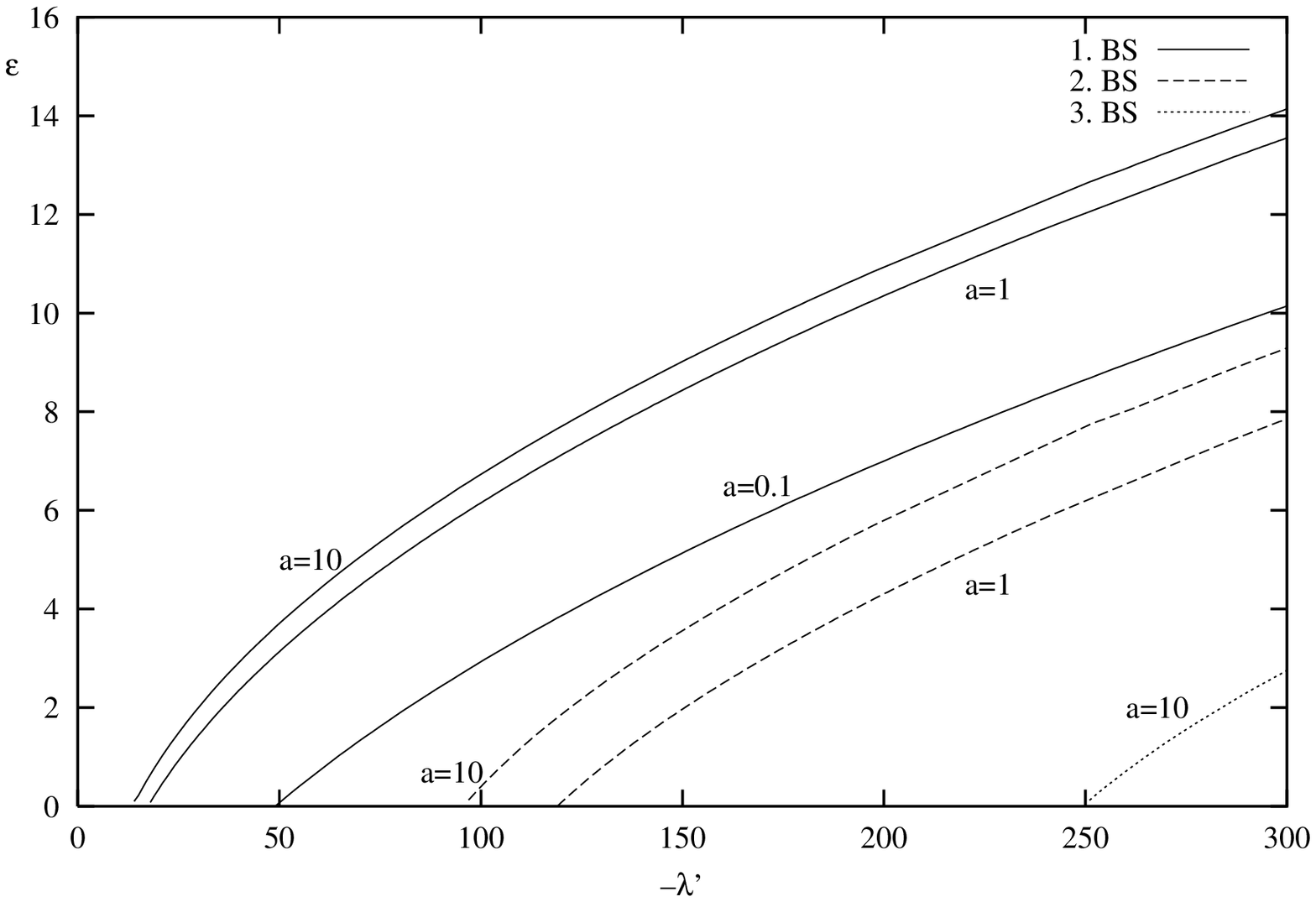}
\end{minipage}
\vskip-10.8cm\hskip-.7cm
  \begin{minipage}[b]{12cm}
\includegraphics[scale=0.9,angle=-90]{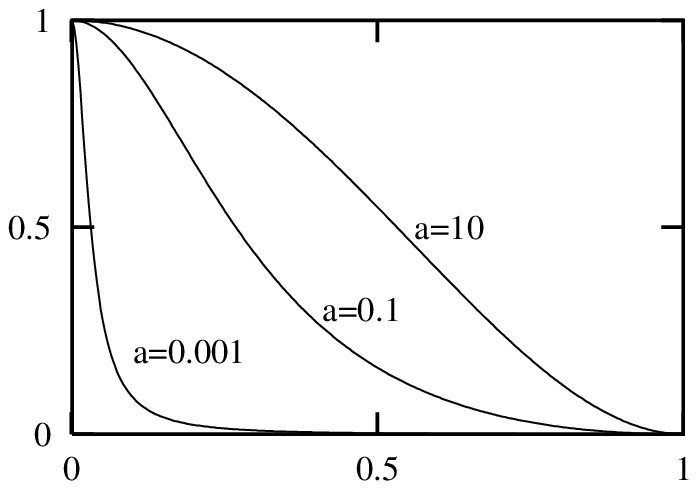}
  \end{minipage}
\vskip7cm
    \caption{Energy of bound states for type A potential and negative
$\lambda^{\prime}$.
(The inset shows  $\varphi(\rho)$ where $V(\rho) =
\lambda^{\prime} \varphi^2(\rho)$)}
    \label{fig:ea}
\end{figure}

\begin{figure}[!htbp]
  \begin{minipage}[t]{18cm}
\hskip-1.5cm
\includegraphics[scale=0.9,angle=-90]{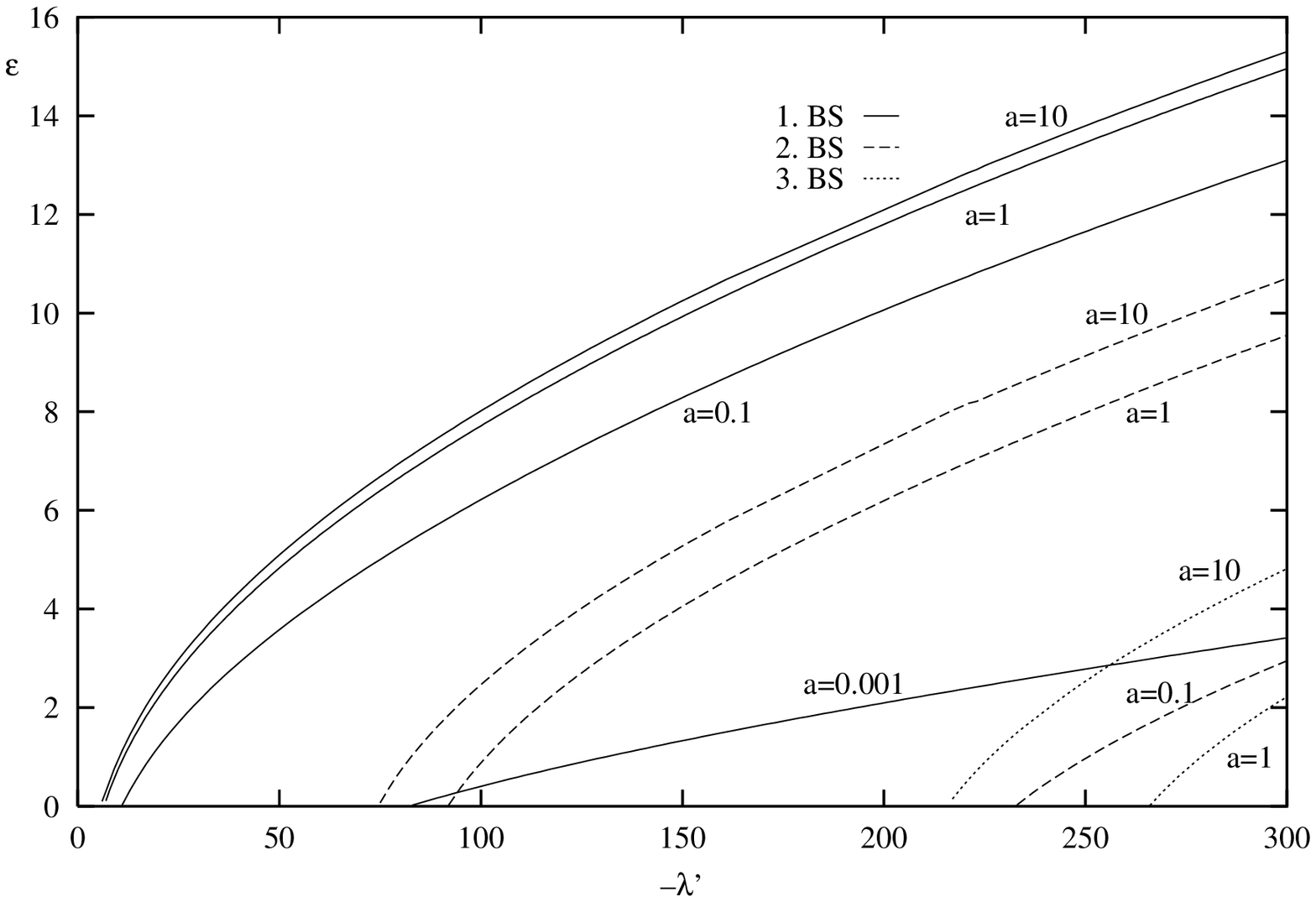}
\end{minipage}
\vskip-10.8cm\hskip-.8cm
  \begin{minipage}[b]{12cm}
\includegraphics[scale=0.85,angle=-90]{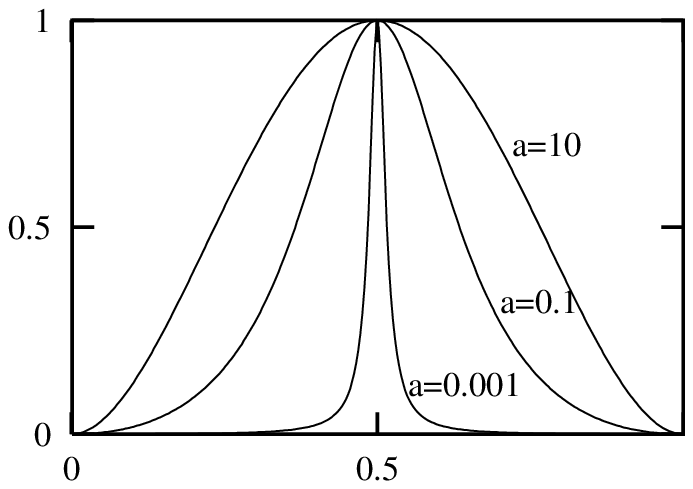}
  \end{minipage}
\vskip7cm
    \caption{Energy of bound states for type B potential and negative
$\lambda^{\prime}$.
(The inset shows  $\varphi(\rho)$ where $V(\rho) =
\lambda^{\prime} \varphi^2(\rho)$)}
    \label{fig:eb}
\end{figure}

The most determining property of the potential $V(r)$ in the attractive case
$\lambda^\prime < 0$ is the number and depth of bound states.  The parameter
dependence of the $l=0$ bound states, determined as zeroes of the $l=0$ Jost
function, Eq.~(\ref{endjost}), is shown in figs.~\ref{fig:ea}
and~\ref{fig:eb}. Number as well as depth of the bound states increases with
increasing $|\lambda^{\prime }|$ and with increasing $a$, too.

\begin{figure}[!htbp]
  \begin{minipage}[t]{18cm}
\hskip-1.5cm\includegraphics[scale=0.9,angle=-90]{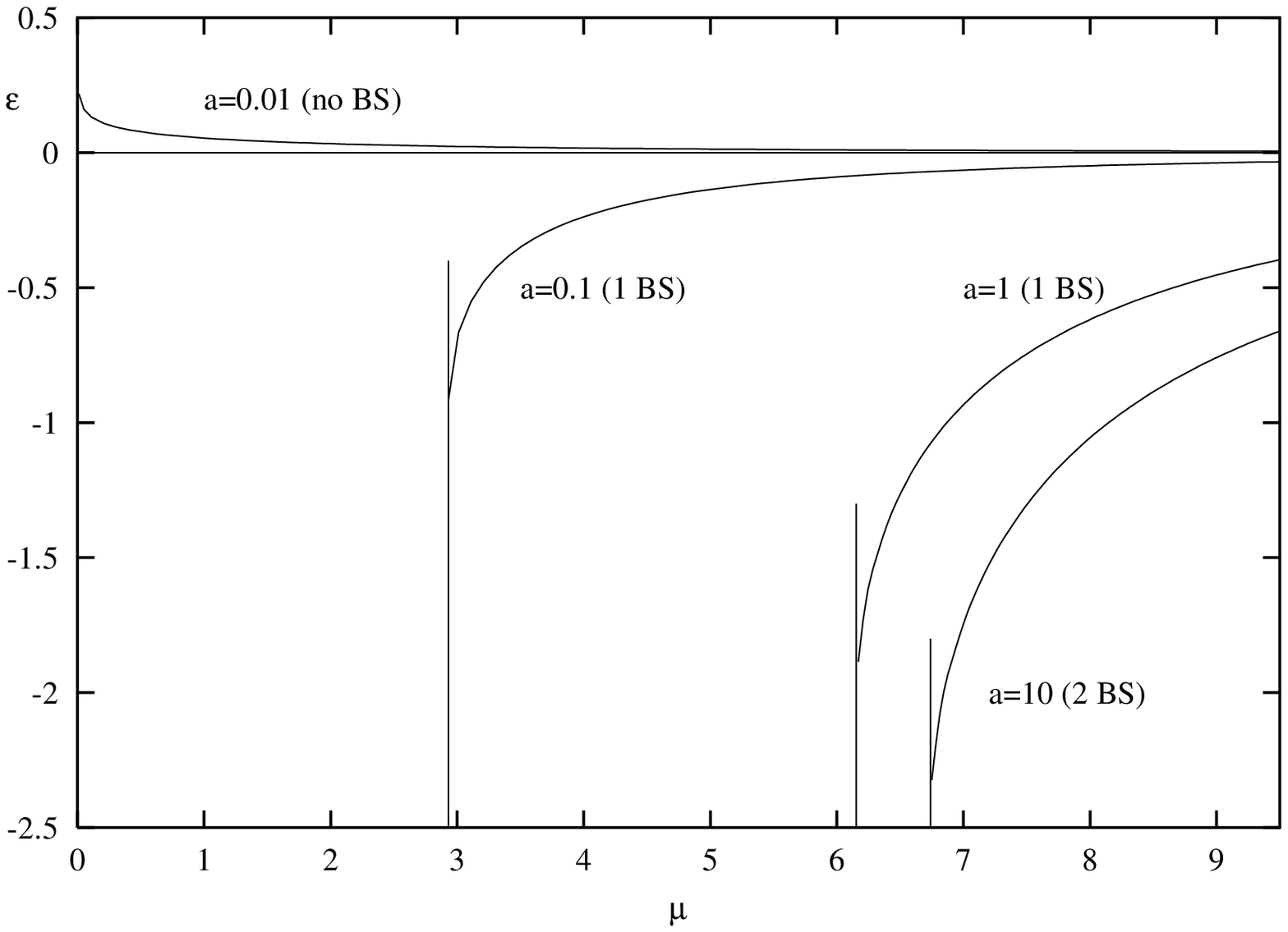}
\end{minipage}
    \caption{Energy for type A potentials of different shapes with
$\lambda^{\prime}=-100$.
      The positions of bound states (BS) at the $\mu$ axis are shown
as vertical
      lines.}
    \label{fig:eea2}
\end{figure}

\begin{figure}[!htbp]
  \begin{minipage}[t]{18cm}
\hskip-1.5cm\includegraphics[scale=0.9,angle=-90]{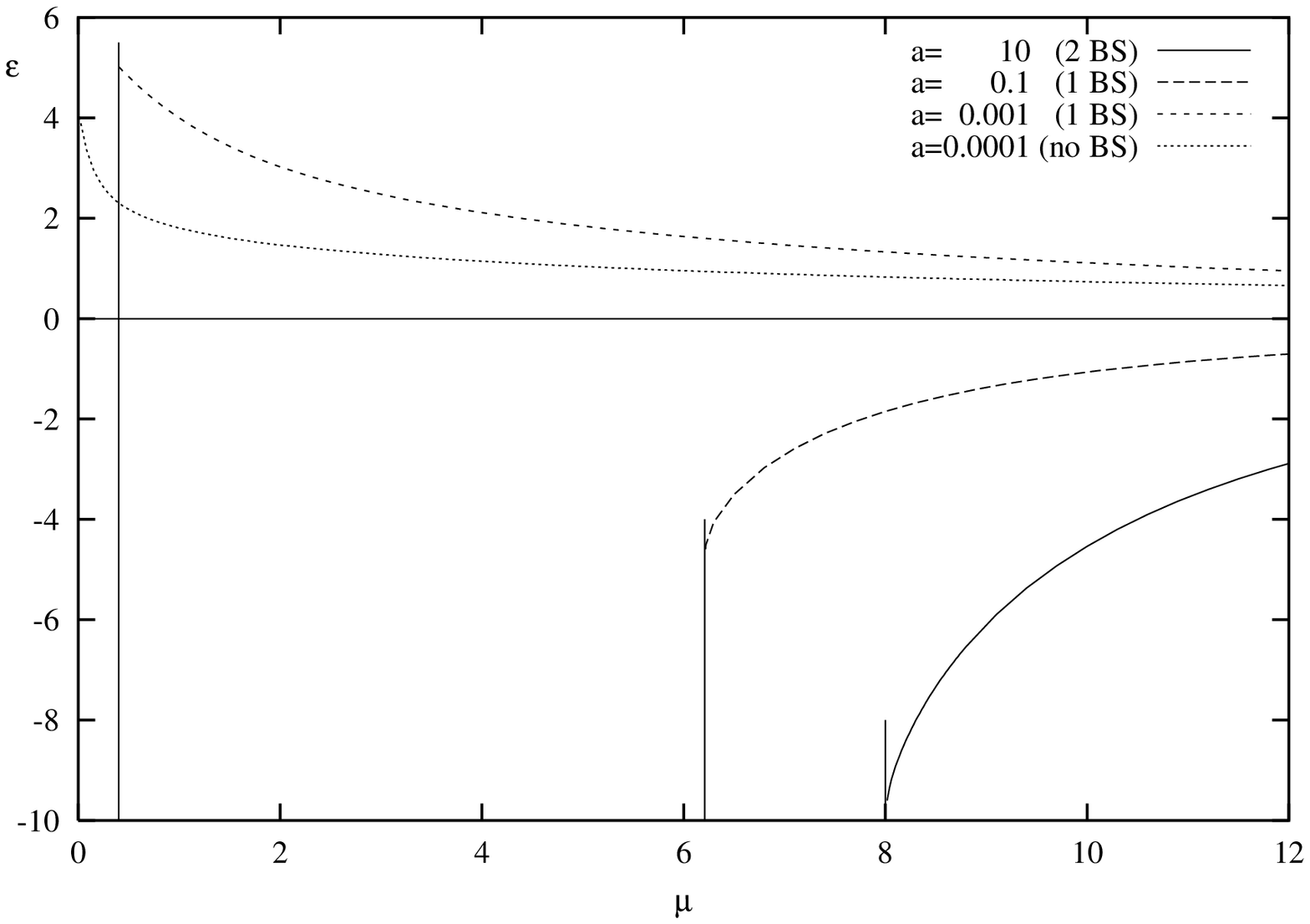}
\end{minipage}
    \caption{Energy for type B potentials of different shapes with
$\lambda^{\prime}=-100$.
      The positions of bound states (BS) at the $\mu$ axis
are shown as vertical
      lines.}
    \label{fig:eeb}
\end{figure}

\begin{figure}[!htbp]
  \begin{minipage}[t]{18cm}
\hskip-1.5cm\includegraphics[scale=0.9,angle=-90]{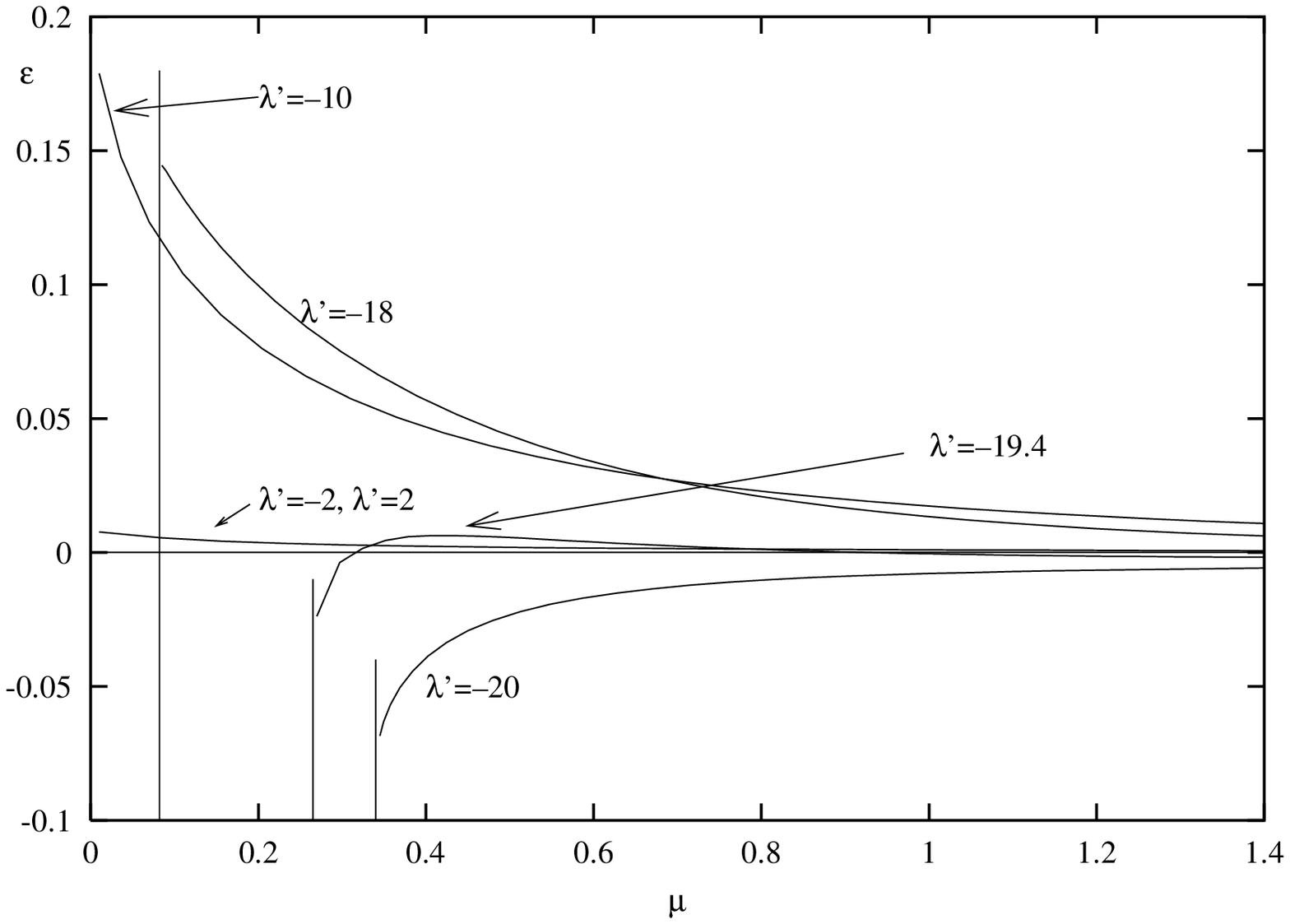}
\end{minipage}
    \caption{Energy for type A potentials of different
magnitudes $\lambda^{\prime}$
     with equal shape parameter $a=1$.
     The positions of bound states existing for
$\lambda^{\prime} = -18, -19.4$ and
     $-20$ at the $\mu$ axis are shown as vertical lines.}
    \label{fig:eea}
\end{figure}

\begin{figure}[!htbp]
  \begin{minipage}[t]{18cm}
\hskip-1.5cm\includegraphics[scale=0.9,angle=-90]{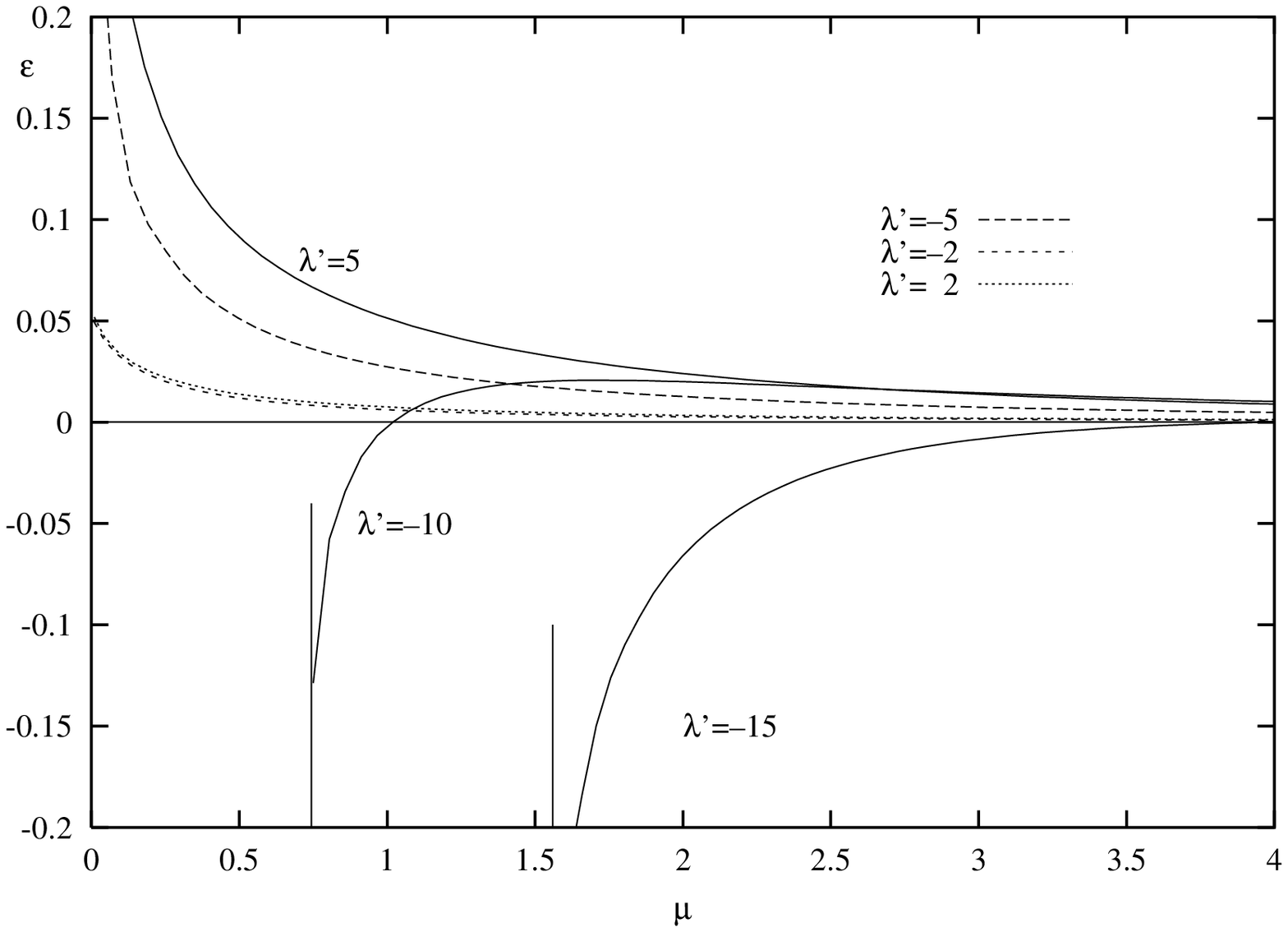}
\end{minipage}
    \caption{Energy for type B potentials of different
magnitudes $\lambda^{\prime}$
     with equal shape parameter $a=1$.  The positions of
      bound states existing for $\lambda^{\prime} = -15$ and
     $-10$ at the $\mu$ axis are shown as vertical lines.}
    \label{fig:eeb2}
\end{figure}

As Eq.~(\ref{efinfin}) shows, the Casimir energy develops an imaginary part
(corresponding to particle creation by the external field) when the
$q$-integration passes a zero of the Jost function. $E_f$ is real as long as
$\mu$ is larger than the depth of the highest bound state.  Therefore we
indicate in our figures~\ref{fig:eea2}--\ref{fig:eeb2} the positions of bound
states on the $\mu$ axis by vertical lines where the plots of the Casimir
energy end.

To show clearly the influence of the form and spectrum of~$V$ on the vacuum
energy, we consider the dependence of the energy for fixed~$\lambda^{\prime }$
and different values of the parameter~$a$, figs.~\ref{fig:eea2}
and~\ref{fig:eeb}, as well as for fixed shape~$a$ and different
depths~$\lambda^{\prime }$ in figs.~\ref{fig:eea} and~\ref{fig:eeb2}.

These figures demonstrate that the vacuum energy is determined by two
competing contributions, the positive contribution of the scattering states
and the negative contribution of the bound states of the potential.

Let us start with a description and interpretation of figs.~\ref{fig:eea2}
and~\ref{fig:eeb}.  For $a$ large, the contribution of the bound state(s) to
the vacuum energy is large enough to overcompensate the contribution of the
scattering states and the energy is negative. At some ``critical value'' of
$a$, the energy of the only remaining bound state is so small that the
positive contributions of the scattering states get the upper hand, and the
vacuum energy is positive. Finally, for $a\to0$ we have normalized
$E_\varphi^{ren} [\Phi =0]=0$, such that at some point $E_\varphi^{ren}$
starts to decrease again with decreasing $a$.

The same features are clearly recovered in figs.~\ref{fig:eea}
and~\ref{fig:eeb2}. For large enough $(-\lambda^{\prime })$ a negative vacuum
energy energy is obtained.  With decreasing $(-\lambda^{\prime })$ the energy
increases and can be positive or negative depending on the parameter $\mu$.
Again, at some point the scattering states dominate, the energy gets positive
and tends to zero as $\lambda^{\prime }$ tends to zero. For positive
$\lambda^{\prime }$ the energy starts to increase again, being virtually
identical for the two values $\lambda^{\prime }= \pm 2$, and it seems that
$E_\varphi^{ren}$ is a monotonically increasing function of $\lambda^{\prime
  }$ for $\lambda^{\prime }\geq 0$. Since we see no qualitative difference
between our two potential types A and B, we expect these features to be
universal.

\begin{figure}[!htbp]
  \begin{minipage}[t]{18cm}
\hskip-1.5cm\includegraphics[scale=0.9,angle=-90]{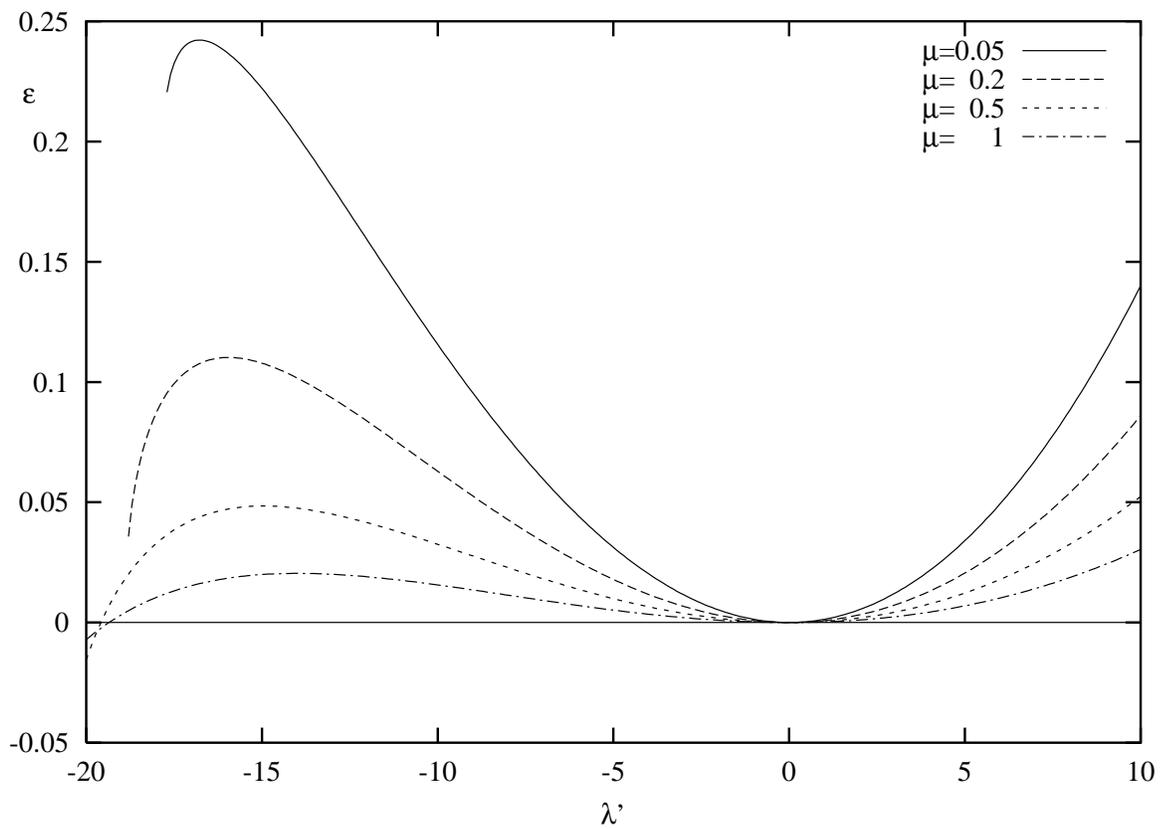}
\end{minipage}
    \caption{Energy for type A potentials for different
magnitudes $\lambda^{\prime}$
      and equal shape parameter $a=1$ as a function of the parameter $\mu$. }
    \label{fig:eea3}
\end{figure}

Given that the dependence of the energy on $\lambda^{\prime }$ is somewhat
nontrivial, let us show this dependence explicitely for type A potentials in
fig.~\ref{fig:eea3}.

As long as no bound states exist, the sign of $\lambda^{\prime }$ is not
important and the energy is nearly symmetric about $\lambda^{\prime }=0$.  The
contribution to the energy of the scattering states does not depend much on
the sign of $\lambda^{\prime }$. The symmetry is disturbed as soon as a bound
state shows up and the energy then decreases with decreasing $\lambda^{\prime
  }$ because the bound state depth as well as the number of bound states
increases.

\section*{Conclusions}
We demonstrated that our method is quite feasible for calculations of the
Casimir energy in realistic 3-dimensional examples for arbitrary
spherical symmetric external
potentials as long as they are smooth and have compact support.

The behavior of the Casimir energy shows some interesting features. The
existence of bound states is a necessary but by no means sufficient condition
for a negative vacuum energy. Even in the presence of bound states the sign of
the energy can depend on the scale parameter $\mu$, i.e the mass of the
quantum field measured in units of the size $R$ of the potential.

For shallow potentials without bound states
(small  $|\lambda^{\prime }|$) the Casimir energy is
nearly invariant under a change of the sign of the potential whereas for deep
attractive potentials the negative contribution of the bound states dominates
the vacuum energy.

On the other hand we found no qualitative differences between a lump-like and
a ring-like classical background configuration.

The method developed here can be used for the calculation of quantum
corrections to classical objects like Skyrmions and others. 

\vspace{12pt}\noindent
KK had been supported by DFG under grant number Bo 1112/4-2.

\section*{Appendix: Analytical continuation of the asymptotic contributions}

The essential formulas for the basic series
$f(s;a;b)$, Eq.~(\ref{funf}), in the calculation are the
following (we use $x=mr$):
\begin{eqnarray}
\sum_{\nu =1/2,3/2,...}^{\infty} \nu^{2n+1} \left(1+\left(\frac{\nu} x
\right) ^2 \right) ^{-s} &=& \frac 1 2 \frac {n! \Gamma (s-n-1)}{\Gamma (s)}
x^{2n+2} \label{continuation1}\\
& &\hspace{-25mm} +(-1)^n 2 \int\limits_0^x d\nu \,\, \frac{\nu^{2n+1} }
{1+e^{2\pi\nu}  } \left( 1-\left( \frac{\nu} x \right)^2 \right) ^{-s}
\nonumber\\
& & \hspace{-25mm} +(-1)^n 2\cos (\pi s)
 \int\limits_x^{\infty} d\nu \,\, \frac{\nu^{2n+1} }
{1+e^{2\pi\nu}  } \left( \left( \frac{\nu} x \right)^2 -1
\right) ^{-s}, \nonumber
\end{eqnarray}
\begin{eqnarray}
\sum_{\nu =1/2,3/2,...}^{\infty} \nu^{2n} \left(1+\left(\frac{\nu} x
\right) ^2 \right) ^{-s} &=& \frac 1 2 \frac {
\Gamma (n+1/2) \Gamma (s-n-1/2)}{\Gamma (s)}
x^{2n+1} \label{continuation2}\\
& &\hspace{-25mm} -(-1)^n 2\sin (\pi s)
 \int\limits_x^{\infty} d\nu \,\, \frac{\nu^{2n} }
{1+e^{2\pi\nu}  } \left( \left( \frac{\nu} x \right)^2 -1
\right) ^{-s}. \nonumber
\end{eqnarray}
Using partial integrations one
can get representations valid for all values of $s$ needed. The explicit
results applied for the calculation of (\ref{as}) are
[we shall use the notation
$f(a;b) = f (-1/2;a;b)$],
\begin{eqnarray}
f(1;1/2) &=&-\frac 1 2 x^2 +\frac 1 {24} \nonumber \\
\frac d {ds}|_{s=-1/2} f(s;1;1/2) &=&
-\frac 1 2 x^2 -2\int_0^{\infty} d\nu
\frac{\nu} {1+e^{2\pi \nu}}
\ln \left|1-\left(\frac{\nu}{x}\right)^2 \right| \nonumber\\
f(1;3/2)&=&\frac{x^2}{2(s+1/2)} + x^2 \ln x \nonumber\\
&&+ x^2
\int_0^\infty d\nu\, \left( \frac{d}{d\nu} \frac{1}{
1+e^{2\pi \nu}} \right)\, \ln \left| \nu^2 -x^2 \right|, \nonumber \\
f(3;5/2)&=&\frac{x^4}{2(s+1/2)}+(\ln x -1/2) x^4 \nonumber\\
&&+\frac{x^4}{2}
 \int_0^\infty d\nu\,
\left[ \frac{d}{d\nu} \left( \frac{1}{\nu} \frac{d}{d\nu} \frac{\nu^2}{
1+e^{2\pi \nu}}\right)  \right]\, \ln \left| \nu^2 - x^2 \right|,
\nonumber \\
f(5;7/2)&=&\frac{x^6}{2(s+1/2)}+(\ln x -3/4) x^6\label{anhfun}\\
&& +\frac{x^6}{8}
 \int_0^\infty d\nu\,
\left[ \frac{d}{d\nu} \left( \frac{1}{\nu} \frac{d}{d\nu}
 \frac{1}{\nu} \frac{d}{d\nu} \frac{\nu^4}{
1+e^{2\pi \nu}}\right)  \right]\, \ln \left| \nu^2 - x^2 \right|,
\nonumber
\end{eqnarray}
which provide the needed expansions about $s=0$ in (\ref{as}).


\end{document}